\newif\ifdoubleblind
\newif\ifacm
	
\ifacm
	\documentclass[sigconf]{acmart}
\else
	\documentclass[conference]{IEEEtran}
\fi
\usepackage{amsmath}

\usepackage{multirow}
\usepackage{tabularx,booktabs}
\usepackage{xspace}
\usepackage{balance}

\usepackage[binary-units=true]{siunitx}
\usepackage{tikz}
\usetikzlibrary{positioning}
\usepackage{mathtools}
\usepackage{amsmath,amsfonts,amssymb}
\usepackage{watermark}

\ifacm
	\usepackage[nolist]{acronym}
\usepackage{booktabs} 
\usepackage{subfig}
\usepackage{balance}

\usepackage{array}
\usepackage{graphicx}
\usepackage{wasysym}
\usepackage{xparse}

\usepackage{makecell}
\newcolumntype{Y}{>{\centering\arraybackslash}X}

\hypersetup{draft}

\renewcommand\footnotetextcopyrightpermission[1]{} 
\else
	\usepackage{graphicx}
\usepackage{epstopdf}
\usepackage{standalone}
\usepackage[nolist ]{acronym}
\usepackage{tcolorbox}
\usepackage{pdfpages}
\usepackage[draft]{hyperref}
\usepackage{pdfcomment}
\usepackage{hologo}

\usepackage{xstring}

\usepackage[utf8]{inputenc}
\usepackage{marvosym}

\usepackage{algorithm}
\usepackage{algpseudocode}

\usepackage{psfrag}
\usepackage{graphicx}

\usepackage{rotating}

\renewcommand{\baselinestretch}{1}

\usepackage[caption=false,font=footnotesize]{subfig}
\usepackage{stfloats}

\usepackage{wasysym}

\fi

\renewcommand\baselinestretch{0.98}

\begin{document}

\newcommand\single{1\textwidth}
\newcommand\double{.48\textwidth}
\newcommand\triple{.32\textwidth}
\newcommand\quarter{.24\textwidth}
\newcommand\singleC{1\columnwidth}
\newcommand\doubleC{.475\columnwidth}
\newcommand\dragon{\ac{DRaGon}\xspace}

\newcommand\red[1]{\colorbox{red}{\textbf{TODO:} #1}}

\newcommand\dortmund{\emph{Dortmund}\xspace}
\newcommand\wuppertal{\emph{Wuppertal}\xspace}
\newcommand\kopenhagen{\emph{Kopenhagen}\xspace}
\newcommand\aarhus{\emph{Aarhus}\xspace}

\newcommand\tikzFig[2]
{
	\begin{tikzpicture}
		\node[draw,minimum height=#2,minimum width=\columnwidth,text width=\columnwidth,pos=0.5]{\LARGE #1};
	\end{tikzpicture}
}

\newcommand{\dummy}[3]
{
	\begin{figure}[b!]  
		\begin{tikzpicture}
		\node[draw,minimum height=6cm,minimum width=\columnwidth,text width=\columnwidth,pos=0.5]{\LARGE #1};
		\end{tikzpicture}
		\caption{#2}
		\label{#3}
	\end{figure}
}

\newcommand\pos{h!tb}

\newcommand{\basicFig}[7]
{
	\begin{figure}[#1]  	
		\vspace{#6}
		\centering		  
		\includegraphics[width=#7\columnwidth]{#2}
		\caption{#3}
		\label{#4}
		\vspace{#5}	
	\end{figure}
}
\newcommand{\fig}[4]{\basicFig{#1}{#2}{#3}{#4}{0cm}{0cm}{1}}

\newcommand\sFig[2]{\begin{subfigure}{#2}\includegraphics[width=\textwidth]{#1}\caption{}\end{subfigure}}
\newcommand\vs{\vspace{-0.3cm}}
\newcommand\vsF{\vspace{-0.4cm}}

\newcommand{\subfig}[3]
{%
	\subfloat[#3]%
	{%
		\includegraphics[width=#2\textwidth]{#1}%
	}%
	\hfill%
}

\newcommand\circled[1] 
{
	\tikz[baseline=(char.base)]
	{
		\node[shape=circle,draw,inner sep=1pt] (char) {#1};
	}\xspace
}
\begin{acronym}
	\acro{5GAA}{5G Automotive Association}
	\acro{DDNS}{Data-Driven Network Simulation}
	\acro{LIMITS}{LIghweight Machine learning for IoT Systems}
	\acro{LIMoSim}{Lightweight ICT-centric Mobility Simulation}
	\acro{LOS}{Line-of-Sight}
	\acro{NLOS}{Non-line-of-Sight}
	\acro{DLOS}{Dynamic Line-of-Sight}
	\acro{REM}{Radio Environmental Map}
	\acro{UAV}{Unmaneed Aerial Vehicle}
	\acro{DRaGon}{Deep RAdio channel modeling from GeOinformatioN}
	\acro{RAIK}{Regional Analysis to Infer KPIs}
	\acro{LTE}{Long Term Evolution}
	\acro{UE}{User Equipment}
	\acro{eNB}{evolved Node B}
	\acro{LIDAR}{Light Detection And Ranging}
	\acro{OSM}{OpenStreetMap}
	\acro{UMa}{Urban Macro}
	\acro{eNB}{evolved Node B}
	\acro{UE}{User Equipment}
	\acro{RSS}{Received Signal Strength}
	\acro{GNSS}{Global Navigation Satellite System}
	\acro{API}{Application Programming Interface}
	\acro{MAE}{Mean Absolute Error}
	\acro{MSE}{Mean Square Error}
	\acro{RMSE}{Root Mean Square Error}
	\acro{NN}{Neural Network}
	\acro{RSRP}{Reference Signal Received Power}
	\acro{EU-DEM}{Digital Elevation Model over Europe}
	\acro{MNO}{Mobile Network Operator}
	\acro{EPS}{Encapsulated Postscript}
	\acro{EIRP}{Equivalent Isotropically Radiated Power}
	\acro{3GPP}{3rd Generation Partnership Project}
	\acro{DNN}{Deep Neural Network}
	\acro{QoS}{Quality of Service}
	\acro{VANET}{Vehicular Ad-hoc Network}
	\acro{RAT}{Radio Access Technology}
	\acro{CUDA}{Compute Unified Device Architecture}
	\acro{PRB}{Physical Resource Block}
	\acro{DPM}{Dominant Path Model}
	\acro{CNN}{Convolutional Neural Network}
	\acro{ReLU}{Rectified Linear Unit}
\end{acronym}

\title{DRaGon: Mining Latent Radio Channel Information from Geographical Data Leveraging Deep Learning}
\author{\IEEEauthorblockN{
\textbf{Benjamin Sliwa}$^1$, 
\textbf{Melina Geis}$^1$, 
\textbf{Caner Bektas}$^1$, 
\textbf{Melisa Lopéz}$^2$, 
\textbf{Preben Mogensen}$^2$, 
\textbf{and Christian Wietfeld}$^1$}
	\IEEEauthorblockA{
		$^1$Communication Networks Institute, TU Dortmund University, 44227 Dortmund, Germany\\
		$^2$Wireless Communication Networks, Aalborg University, 9220 Aalborg, Denmark\\
		e-mail: $^1\{$firstname.lastname$\}$@tu-dortmund.de $^2\{$mll, pm$\}$@es.aau.dk
	}
}

\maketitle	
\begin{abstract}
	%
	%
	Radio channel modeling is one of the most fundamental aspects in the process of designing, optimizing, and simulating wireless communication networks. In this field, long-established approaches such as analytical channel models and ray tracing techniques represent the de-facto standard methodologies.
	%
	%
	However, as demonstrated by recent results, there remains an untapped potential to innovate this research field by enriching model-based approaches with machine learning techniques.
	%
	%
	In this paper, we present \dragon as a novel machine learning-enabled method for automatic generation of \acp{REM} from geographical data. For achieving accurate path loss prediction results, \dragon combines determining features extracted from a three-dimensional model of the radio propagation environment with raw images of the receiver area within a deep learning model.
	%
	%
	In a comprehensive performance evaluation and validation campaign, we compare the accuracy of the proposed approach with real world measurements, ray tracing analyses, and well-known channel models.
	%
	%
	It is found that the combination of expert knowledge from the communications domain and the data analysis capabilities of deep learning allows to achieve a significantly higher prediction accuracy than the reference methods.
\end{abstract}

\begin{tikzpicture}[remember picture, overlay]
\node[below=5mm of current page.north, text width=20cm,font=\sffamily\footnotesize,align=center] {Accepted for presentation in: 2022 IEEE Wireless Communications and Networking Conference (WCNC)\vspace{0.3cm}\\\pdfcomment[color=yellow,icon=Note]{
@InProceedings\{Sliwa2022a,\\
	Author = \{Benjamin Sliwa, Melina Geis, Caner Bektas, Melisa Lopéz, Preben Mogensen, and Christian Wietfeld\},\\
	Title = \{\{DRaGon\}: \{M\}ining latent radio channel information from geographical data leveraging deep learning\},\\
	Booktitle = \{2022 IEEE Wireless Communications and Networking Conference (WCNC)\},\\
	Year = \{2022\},\\
	Address = \{Austin, Texas, USA\},\\
	Month = \{Apr\},\\
\}
}};
\node[above=5mm of current page.south, text width=15cm,font=\sffamily\footnotesize] {2022~IEEE. Personal use of this material is permitted. Permission from IEEE must be obtained for all other uses, including reprinting/republishing this material for advertising or promotional purposes, collecting new collected works for resale or redistribution to servers or lists, or reuse of any copyrighted component of this work in other works.};
\end{tikzpicture}

\section{Introduction}

%
%
The ability to determine the received signal strength for a given sender and receiver pair is one of the most fundamental aspects for planning \cite{Taufique/etal/2017a} and simulating \cite{Cavalcanti/etal/2018a} wireless networks. Moreover, emerging mobile communication paradigms such as anticipatory networking \cite{Bui/etal/2017a, Sliwa/etal/2021b} explicitly build upon using a priori context knowledge for proactive network optimization.

%
%
However, the existing model-based approaches for path loss determination heavily rely on \emph{abstractions} and simplifications of the radio propagation environment, which limits their significance for \emph{concrete} real world scenarios.
%
%
Ray tracing \cite{Yun/Iskander/2015a} aims to overcome these limitations through explicit modeling of the radio propagation environment. However, in many practical scenarios, the required high resolution data about shapes and materials of the obstacles might not be available, inherently implying a significant degradation of the achievable modeling accuracy \cite{Thrane/etal/2020b}.
%
%
Fueled by the availability of computation power, data sets, and algorithms, machine learning has started to become an integral part of all areas related to wireless networking \cite{Ali/etal/2020a}. Due to its inherent strength of learning closed system descriptions of complex processes from measurable features, it has also become a promising method for solving classical communication problems such as network simulation \cite{Sliwa/Wietfeld/2019a} and radio channel modeling \cite{Wang/etal/2020b}.

%
%
\begin{figure}[]  	
	\vspace{0cm}
	\centering		  
	\includegraphics[width=1.0\columnwidth]{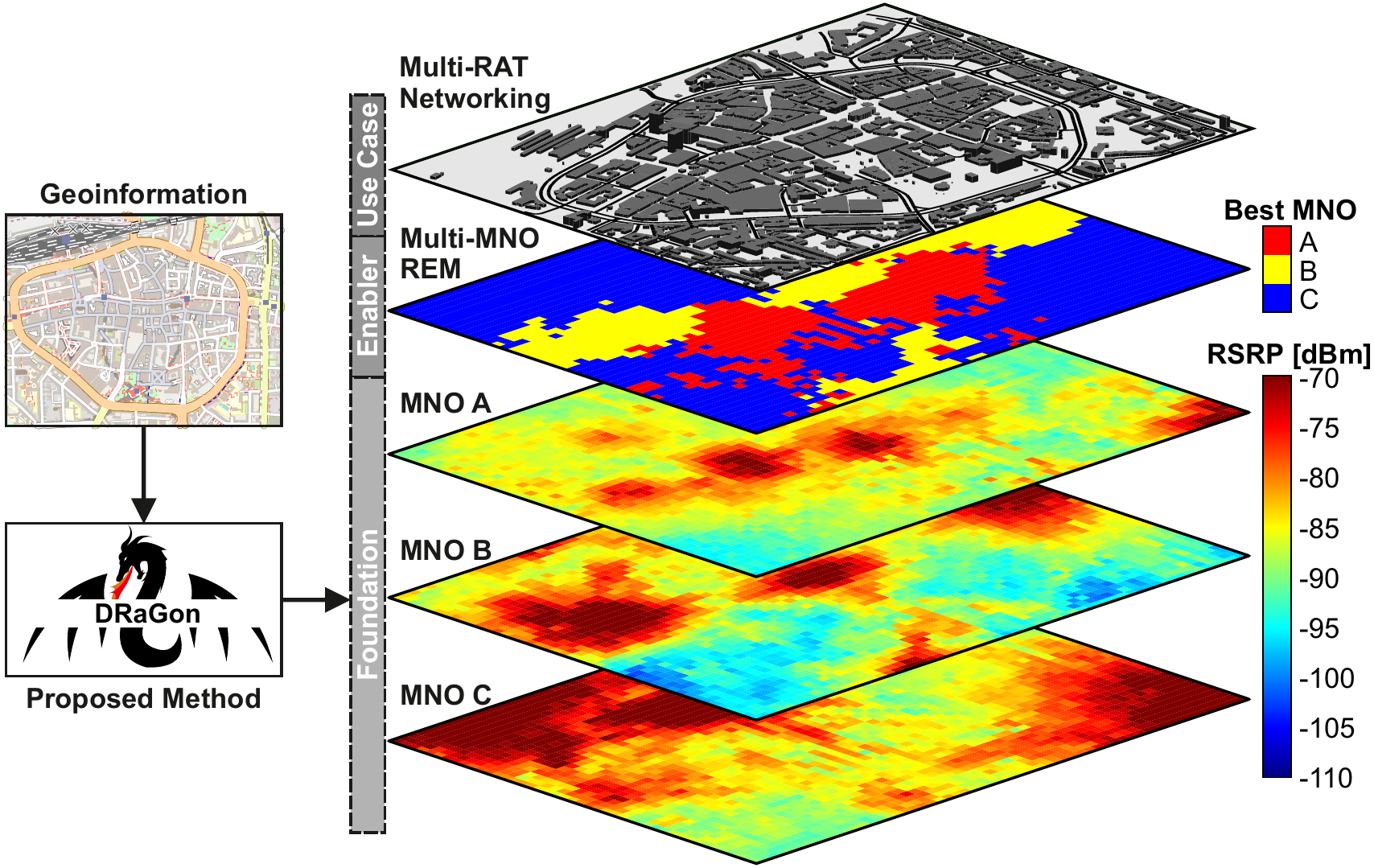}
	\caption{Example application of the proposed \dragon method for generating RSRP \acp{REM} for three different \ac{MNO}. The acquired radio channel knowledge is aggregated in a multi-\ac{MNO} map that provides the required information for performing dynamic \ac{MNO} selection at the application layer (Map data: $\copyright$ OpenStreetMap Contributors, CC BY-SA).}
	\label{fig:scenario}
	\vspace{-0.5cm}	
\end{figure}
%
%
In this paper, we present the novel \dragon method for extracting latent radio channel information from widely available geographical data that builds upon the well-known strengths of deep learning-based \cite{Goodfellow/etal/2016a} image analysis.
%
%
Following the assumption that similar ``looking'' environments will show similar radio propagation characteristics, \dragon not only relies on explicitly extracted features (e.g., the number of building penetrations on the direct path between sender and receiver) but also incorporates top and side view images of the receiver environment within the prediction process.
%
%
Fig.~\ref{fig:scenario} shows an example for the utilization of the proposed method in the context of multi-\ac{RAT} networking. Hereby, \dragon automatically transforms the geoinformation input data into \ac{MNO}-specific \acp{REM} of the \ac{RSRP}. This information provides the foundation for a multi-\ac{MNO} \ac{REM} that allows to dynamically determine the serving \ac{MNO} based on the current user position.

%
%
The remainder of the paper is structured as follows. After discussing the related work in Sec.~\ref{sec:related_work}, we present the novel \dragon method in Sec.~\ref{sec:approach}. Afterwards, an overview of the methodological aspects is given in Sec.~\ref{sec:methods}. Finally, detailed results of the achievable modeling accuracy and the generalizability of the proposed method  are provided in Sec.~\ref{sec:results}.

\section{Related Work} \label{sec:related_work}

%
%
\textbf{Anticipatory mobile networking} \cite{Bui/etal/2017a} is novel paradigm in wireless communications that focuses on the exploitation of \emph{context} information for proactive network optimization. 
%
%
According to a recent white paper \cite{5GAA/2020a} of the \ac{5GAA}, this form of predictive \ac{QoS} optimization will be one of the key enablers for future connected and autonomous driving.
%
%
However, while the vehicles are able to measure the network quality at their \emph{current} locations, they are not able to determine the corresponding information at the \emph{future} locations along their trajectories in advance. A promising data-driven solution approach for closing these gaps is the utilization of \acp{REM} \cite{Poegel/Wolf/2015a, Kliks/etal/2020a} that map geospatial locations to corresponding previously acquired network context measurements.

%
%
\textbf{Radio channel modeling} --- in addition to purely data-driven approaches such as mobile crowdsensing --- is an important method for constructing the requried \acp{REM}. 
%
%
An overview of the different classes of channel and propagation models for vehicular communications is given by Viriyasitavat et al. in \cite{Viriyasitavat/etal/2015a}. 
In addition to simple analytical models --- that do not integrate explicit environment modeling --- such as Friis and Two-ray ground reflection, different methods have been proposed to increase the significance of the results for \emph{concrete} evaluation scenarios.
%
%
\emph{Empirical} models, such as the \ac{3GPP} \ac{UMa} \cite{3GPP/2019d} model, rely on real world measurement data for specific radio propagation \emph{prototypes} (e.g., urban, suburban, and rural) and distinguish between \ac{LOS} and \ac{NLOS} channel characteristics based on a distance-dependent probability function.
%
%
\emph{Deterministic} channel models \cite{Sommer/etal/2014a} utilize environmental models for computing the building penetrations in order to derive the total path loss by superpositioning the effects of the obstructed path and the non-obstructed path.
%
%
Ultimately, computationally expensive \emph{ray tracing} techniques \cite{Yun/Iskander/2015a} utilize high resolution environmental models for tracing the behavior of the emitted rays based on models for the physical processes such as reflection and refraction.
%
%
In a comprehensive empirical analysis \cite{Cavalcanti/etal/2018a}, Cavalcanti et al. analyzed 283 vehicular networking papers from top-tier conferences and journals. While only 83 of the 214 papers that made use of radio propagation models specified which model was used, the analysis revealed a clear dominance of simple analytical models such as Nakagami, Two-ray ground, Friis, and Rayleigh fading.

%
%
\textbf{Machine learning} has started to penetrate all areas related to wireless communications. Consequently, different research works have given a glimpse at the hidden potential of machine learning-enabled radio propagation modeling.
%
%
A comprehensive summary of different disciplines, models, and applications of machine learning in wireless networking is provided by Wang et al. in \cite{Wang/etal/2020a}.
%
%
In \cite{Enami/etal/2018a}, Enami et al. present \ac{RAIK}, a method for \ac{RSRP} prediction through determining the most suitable path loss exponent of a given channel model. For this purpose, the authors use a \ac{LIDAR} environmental model from which statistical features, such as the percentage of the areas covered by buildings, is extracted.
%
%
A related approach is proposed by Masood et al. in \cite{Masood/etal/2019a}. By combining \ac{eNB}- and \ac{UE}-specific features with geographic information, the authors are able to achieve an \ac{RSS} \ac{RMSE} of  \SI{6.2}{\decibel} in comparison to a ray tracing setup serving as the ground truth.
%
%
The authors of \cite{MorochoCayamcela/etal/2020a}, utilize aerial images of the environment between sender and receiver for classifying the radio channel into urban, suburban, and rural prototypes. For each of these, a specific path loss model is then used for performing predictions.
%
%
Thrane et al. propose an even more consequent approach for the utilization of two-dimensional geoinformation in \cite{Thrane/etal/2020a}. Hereby, raw aerial \emph{images} of the receiver environment are utilized as input features for a deep neural network that learns an environment-dependent correction offset of a path loss model. This methodological approach represents the foundation for the novel \dragon method.

\section{Mining Latent Radio Channel Information from Geographical Data with \ac{DRaGon}} \label{sec:approach}

%
%
\textbf{Problem statement}: Our overall goal is to determine the \ac{RSRP} at a specific receiver position  $\mathbf{p}_{\text{RX}}$ given the transmitter position $\mathbf{p}_{\text{TX}}$. According to the \ac{3GPP} standardization \cite{3GPP/2020a}, the \ac{RSRP} is calculated as
%
%
\begin{equation}
	\text{RSRP} = P_{\text{RX}} - 10 \log_{10}\left(N_{\text{PRB}} \cdot N_{\text{SC}} \right)
\end{equation}
whereas $P_{\text{RX}}$ represents the received signal strength, $N_{\text{PRB}}$ is the number of \acp{PRB}, and $N_{\text{SC}}$ is the number of subcarriers. $N_{\text{PRB}}$ can be derived from the channel bandwidth $B$ (e.g., 100~\acp{PRB} are available for \SI{20}{\mega\hertz} cells) and $N_{\text{SC}}$ is fixed to $12$ for conventional \ac{LTE} systems.

%
%
However, as $P_{\text{RX}}$ is unknown, it is substituted with a generic link budget term $P_{\text{RX}} = P_{\text{TX}} - L + \Delta L$ that allows us to formulate
%
%
\begin{equation} \label{eq:dragon_formula}
	\text{RSRP} = \underbrace{P_{\text{TX}} - 10 \log_{10}\left(N_{\text{PRB}} \cdot N_{\text{SC}} \right) }_{\mathclap{\substack{\text{Properties of the} \\ \text{communication system}}}}
	 - \underbrace{\vphantom{P_{\text{TX}} - 10 \log_{10}\left(N_{\text{PRB}} \cdot N_{\text{SC}} \right)} L}_{\mathclap{\substack{\text{Channel} \\ \text{model}}}}
	 + \underbrace{
	 	\vphantom{P_{\text{TX}} - 10 \log_{10}\left(N_{\text{PRB}} \cdot N_{\text{SC}} \right)} 
	 	\Delta L}_{\mathclap{\substack{\text{ML-based} \\ \text{correction}}}}  .
\end{equation}
%
%
Hereby, $P_{\text{TX}}$ represents an \ac{EIRP} description of the transmitter antenna that aggregates transmission power, antenna gains, and coupling losses. With $L$ being an analytical path loss estimation --- we utilize a \ac{3GPP} \ac{UMa}~B \cite{3GPP/2019d} model for this task --- with respect to $\mathbf{p}_{\text{TX}}$ and $\mathbf{p}_{\text{RX}}$, machine learning techniques are leveraged to learn a correction offset $\Delta L$ using geographical features.
%
%
\begin{figure*}[]  	
	\vspace{0cm}
	\centering		  
	\includegraphics[width=1.0\textwidth]{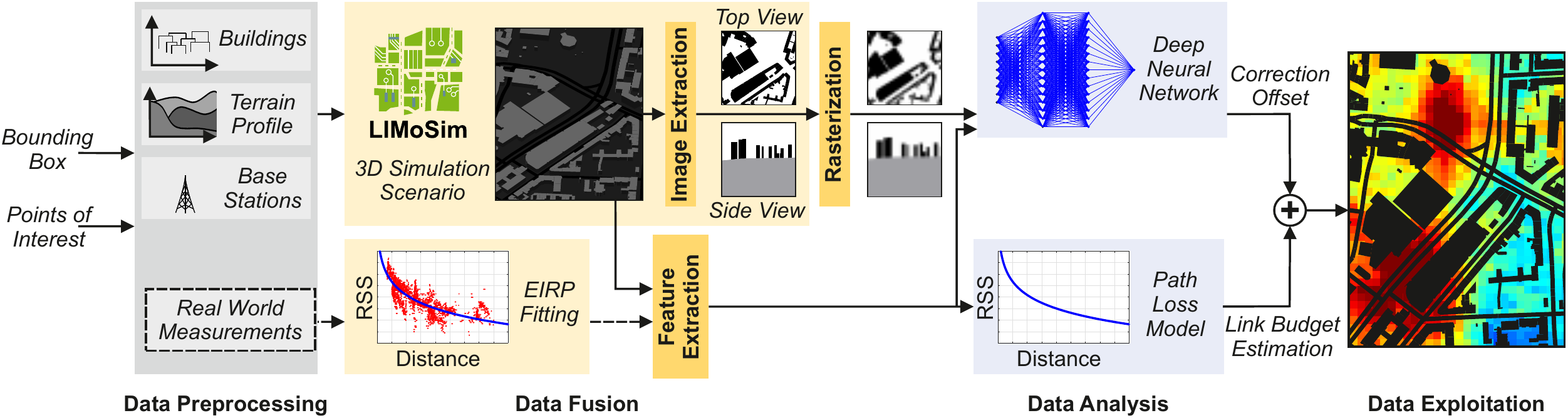}
	\vspace{-0.5cm}	
	\caption{Overall system architecture model of the proposed \ac{DRaGon} signal strength prediction method.}
	\label{fig:architecture}
	\vspace{-0.5cm}	
\end{figure*}
In the following paragraphs, a detailed description of the data processing pipeline is provided. A schematic illustration of the system architecture model of \dragon is shown in Fig.~\ref{fig:architecture}.

%
%
\subsection{Data Preprocessing and Augmentation} \label{sec:data_preprocessing}

%
%
The goal of the initial preprocessing phase is to prepare the input data such that a three-dimensional model of the radio propagation environment --- which is capable of extracting the determining features --- can be constructed.

%
%
\textbf{Radio propagation environment}: 
%
%
Publicly available \ac{OSM} geoinformation is utilized as the main data source for building the environmental model for the given bounding box of the evaluation scenario.
%
%
While the outlines of the buildings --- the dominant sources of signal attenuation --- are well captured within available \ac{OSM} data, height information is not very well represented. As an example, from the 9630 buildings of the urban subset of the \dortmund scenario (see Sec.~\ref{sec:methods}), only 78 are annotated with corresponding height information.
%
In order to close these gaps, we utilize available \ac{LIDAR} data for the German and Danish application scenarios.
%
%
For each \ac{OSM} building, the building height is calibrated with the corresponding data from the \ac{LIDAR} data set whereas the matching is performed based on the respective bounding boxes.

%
%
In addition to the building-related radio propagation effects, ground reflections also have an impact on the \ac{RSRP} and need to be considered. For this purpose, \dragon incorporates data from the \ac{EU-DEM} \cite{Bashfield/Keim/2011a} terrain profile that provides elevation data with \SI{25}{\meter} cell resolution and a vertical accuracy of \SI{7}{\meter} \ac{RMSE}.

\begin{figure}[b]  	
	\vspace{-0.6cm}	
	\centering		  
	\includegraphics[width=1.0\columnwidth]{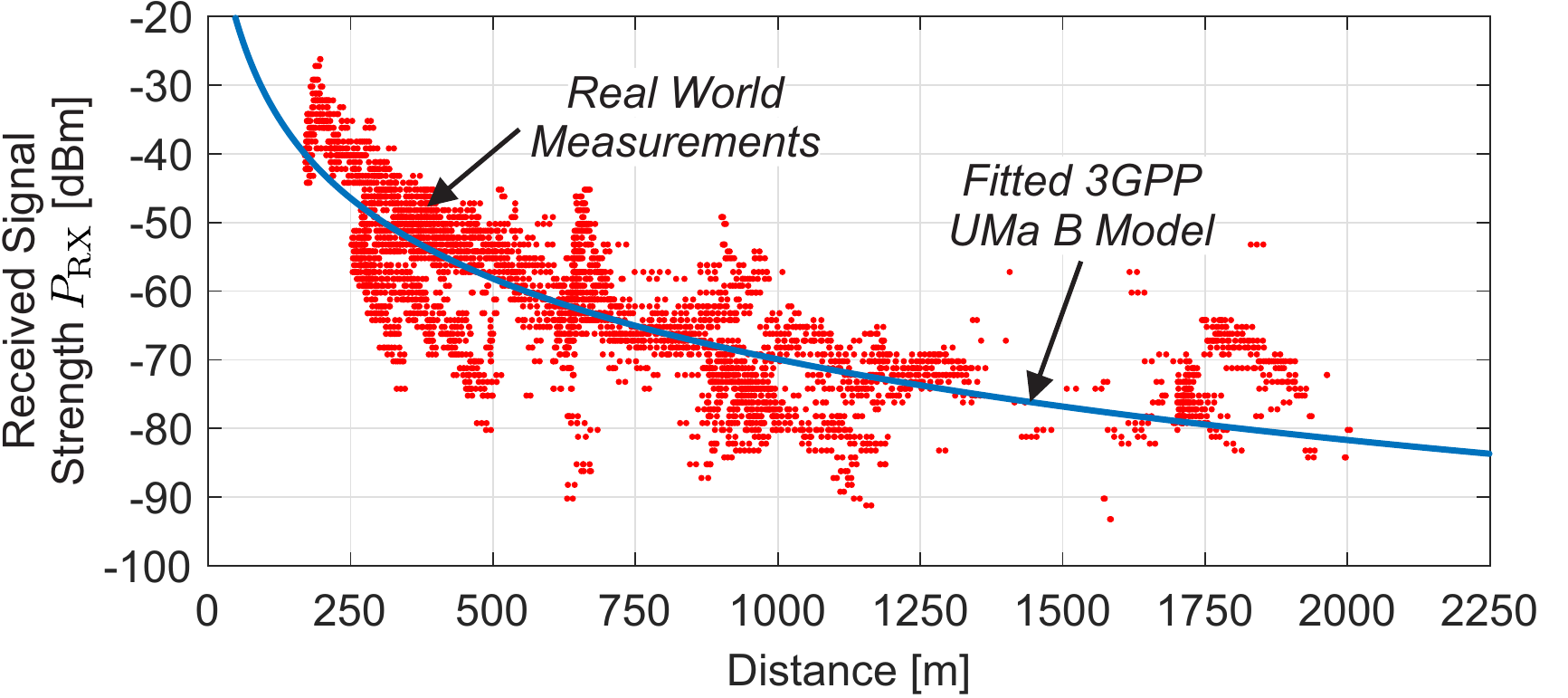}
	\vspace{-0.6cm}	
	\caption{Example for the estimation of the cell-specific \ac{EIRP} $\tilde{P}_{\text{TX}}$ by fitting available real world measurements to an \ac{UMa}~B channel model.}
	\label{fig:uma_fitting}
	\vspace{0cm}	
\end{figure}

%
%
\textbf{Transmission power estimation:}
%
%
For the application of the proposed approach, a major challenge is that information about the transmission power $P_{\text{TX}}$ of the base stations --- required for Eq.~\ref{eq:dragon_formula} --- is typically not publicly available. In order to close this gap, we perform a transmission power estimation step. Hereby, we aim to find the best fit of an \ac{UMa}~B path loss model \cite{3GPP/2019d} to existing real world measurements (see Sec.~\ref{sec:methods}) of the received signal strength $P_{\text{RX}}$. An example for the fitting process is shown in Fig.~\ref{fig:uma_fitting}.
For each cell, the estimated \ac{EIRP} $\tilde{P}_{\text{TX}}$ is determined by minimizing the \ac{MSE} of the $N$ ground truth measurements and the corresponding model predictions $L$ using the objective function 
%
%
\begin{equation}
	\min_{\tilde{P}_{\text{TX}}} \left( \frac{1}{N} \sum_{i=1}^{N} \left( P_{\text{RX},i} - \tilde{P}_{\text{TX}} + L_{i} \right)^2  \right) .
\end{equation}
%
%

%
%
\subsection{Data Fusion and Feature Extraction}

%
%
The augmented data is then utilized for setting up a \ac{LIMoSim} \cite{Sliwa/etal/2019c} scenario. Although \ac{LIMoSim} is actually a simulation framework for ground-based and aerial vehicular mobility, we utilize this unconvential method because of its rich environmental data aggregation capabilities that allow us to fuse the different data sources into a single methodological setup for further processing. For each receiver position $\mathbf{p}_{\text{RX}}$, different types of features are computed:

%
%
\textbf{Receiver environment images}: As one of its unique features, \ac{LIMoSim} provides a dedicated \ac{EPS} rendering engine capable of exporting vector screenshots of defined regions within the simulation scenario. \dragon utilizes this approach for incorporating the raw top and side view images (examples are shown in Fig.~\ref{fig:architecture}) of the receiver environment into the machine learning process.
%
%
Each top view image covers an area of $\SI{300}{\meter} \times \SI{300}{\meter}$ with centered receiver position $\mathbf{p}_{\text{RX}}$. A normalization of the image rotation is performed such that the right axis always points towards the base station. This parameterization is chosen with respect to previous work \cite{Thrane/etal/2020a}, which also provides formal description of the export procedure.
%
%
For generating the side view images, a simple direct path ray tracing is performed for determining the intersections points with terrain and buildings. Within these images, the receiver is vertically centered at the left edge and different colors are chosen for buildings (\emph{black}) and terrain (\emph{gray}).
%
%
Finally, the vector images are converted into $64 \times 64$-rasterized representations $\mathbf{I}_{\text{top}}$ and $\mathbf{I}_{\text{side}}$ to allow their utilization as \ac{NN} input features.

%
%
\textbf{Numerical features}: In addition to the images, several numerical features are extracted from the \ac{LIMoSim} scenario. The aggregated feature vector $\mathbf{x}$ aggregates information from different logical domains:
%
%
\begin{itemize}
	%
	%
	\item \emph{Relative locations}: Absolute latitudinal and longitudinal distances $\Delta_{\text{lon}}$ and $\Delta_{\text{lat}}$ of receiver and transmitter
	%
	%
	\item \emph{Direct path:} Distance $d$ between receiver and transmitter, Number of building intersections $N_{\text{obs}}$, Indoor distance $d_{\text{obs}}$, Number of terrain intersections $N_{\text{ter}}$, Terrain distance $d_{\text{ter}}$ 
	%
	%
	\item \emph{Communication system}: Bandwidth $B$, Carrier frequency $f$, Transmission power estimation $\tilde{P}_{\text{TX}}$
\end{itemize}

%
%
\subsection{Deep Learning-Enabled RSRP Prediction}

%
%
\begin{figure}[]  	
	\centering		  
	\includegraphics[width=1.0\columnwidth]{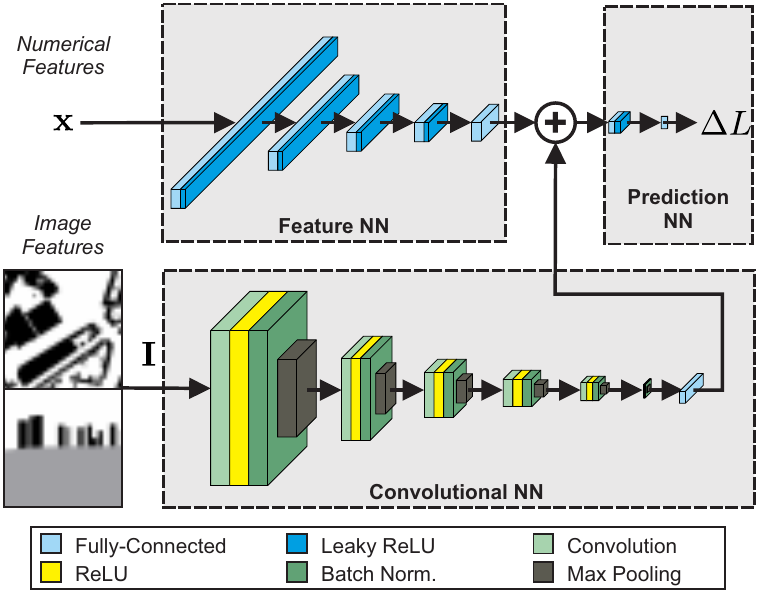}
	\vspace{-0.5cm}	
	\caption{Schematic illustration of the final deep neural network architecture.}
	\label{fig:ann}
	\vspace{-0.5cm}	
\end{figure}
As illustrated in Fig.~\ref{fig:ann}, the derived features are fed into a deep neural network that consists of three sub-\acp{NN}, which handle dedicated processing tasks.
%
%
The \emph{feature \ac{NN}} is utilized for processing the numerical features $\mathbf{x}$ using a sequence of processing blocks that perform linear transformations of the input data using the learned connection weights, followed by element-wise \ac{ReLU} activation and batch normalization.
%
%
In addition, the vertically concatenated images $\mathbf{I} = \begin{bmatrix} \mathbf{I}_{\text{top}}^\intercal, \mathbf{I}_{\text{side}}^\intercal \end{bmatrix}^\intercal$ are handled by the \emph{convolutional \ac{NN}}. Each processing block of the pipeline consists of a convolution layer with zero padding followed by \ac{ReLU} activation, batch normalization and $2 \times 2$ max pooling. The final flattening of the two-dimensional input to the one-dimensional output is performed by the final linear layer.
%
%
The derivation of the target variable $\Delta L$ is then performed using the \emph{prediction \ac{NN}}.
%
%
Finally, the \ac{RSRP} is predicted using Eq.~\ref{eq:dragon_formula}.

\section{Methodology} \label{sec:methods}

%
%
\textbf{Evaluation scenarios:} 
For the performance evaluation, we consider real world measurements from different data sources:
%
%
\begin{itemize}
	%
	%
	\item Vehicular measurements from the German city \dortmund \cite{Sliwa/Wietfeld/2019a} in campus, urban, suburban, and highway environments with three \acp{MNO} (68314 data samples)
	%
	%
	\item Vehicular measurements from the German city \wuppertal \cite{Sliwa/etal/2021b} in the networks of three \acp{MNO} (41113 data samples)  
	%
	%
	\item Vehicular measurements from the Danish city \kopenhagen \cite{Thrane/etal/2020b} in a campus environment (57586 data samples)
	%
	%
	\item \ac{UAV} measurements from the Danish city \aarhus \cite{Lopez/etal/2019a} (268534 data samples)
\end{itemize}

%
%
\textbf{Validation methods:} In addition to the real world measurements, multiple validation methods are used as references:
%
%
\begin{itemize}
	%
	%
	\item \emph{Conventional channel} models such as Friis, Nakagami ($m=2$), and Two-Ray Ground
	%
	%
	\item \emph{Empirical channel} modeling with \ac{3GPP} \ac{UMa}~B \cite{3GPP/2019d} and WINNER II C2 \ac{NLOS}
	%
	%
	\item \emph{Environment-aware} models with Obstacle shadowing \cite{Sommer/etal/2011a} and \texttt{Altair WinProp} ray tracing 
\end{itemize}

%
%
The machine learning evaluations are performed with \texttt{PyTorch} and accelerated with an Nvidia Tesla K40M with 2880 \ac{CUDA} cores.
%
%
For the training, the overall data set $\mathcal{D}$ is split into \SI{80}{\percent} training data $\mathcal{D}_{\text{train}}$, \SI{10}{\percent} test data $\mathcal{D}_{\text{test}}$, and \SI{10}{\percent} validation data $\mathcal{D}_{\text{val}}$.
%
%
For increasing the significance and repeatability of our results, our methodological setup is provided in an Open Source manner\footnote{The source code of the proposed \dragon method is available at \\ \url{https://github.com/melgeis/DRaGon}}.
%
%
A summary of the resulting parameterization of is shown in Tab.~\ref{tab:parameters}.

\vspace{-0.2cm}
%
%
\begin{table}[ht]
	\centering
	\caption{Final Configuration of the DRaGon Hyperparameters}
	\vspace{-0.2cm}
	\begin{tabularx}{\columnwidth}{XX}
		\toprule
		\textbf{Hyperparameter} & \textbf{Value} \\

		\midrule
	
		Stride & 1 \\
		Dilation & 1 \\
		Batch Size & 128 \\
		Zero Padding & 3 \\
		CNN Filters & [32, 16, 16, 16, 10, 1] \\
		Max Pooling & [2, 2, 2, 2, 2, 2] \\
		Kernels & [5, 3, 3, 3, 3, 2] \\
		Feature NN & [256, 128, 64, 32] \\
		Prediction NN & [16] \\
		Learning Rate & 0.001 \\
		Weight Decay & 0.0005 \\ 
		Optimizer & Adam \\
			
		\bottomrule
		
	\end{tabularx}
	\label{tab:parameters}
\end{table}

\vspace{-0.7cm}
\section{Results} \label{sec:results}
\vspace{-0.1cm}

%
%
\begin{figure}[]  	
	\vspace{0cm}
	\centering		  
	\includegraphics[width=1.0\columnwidth]{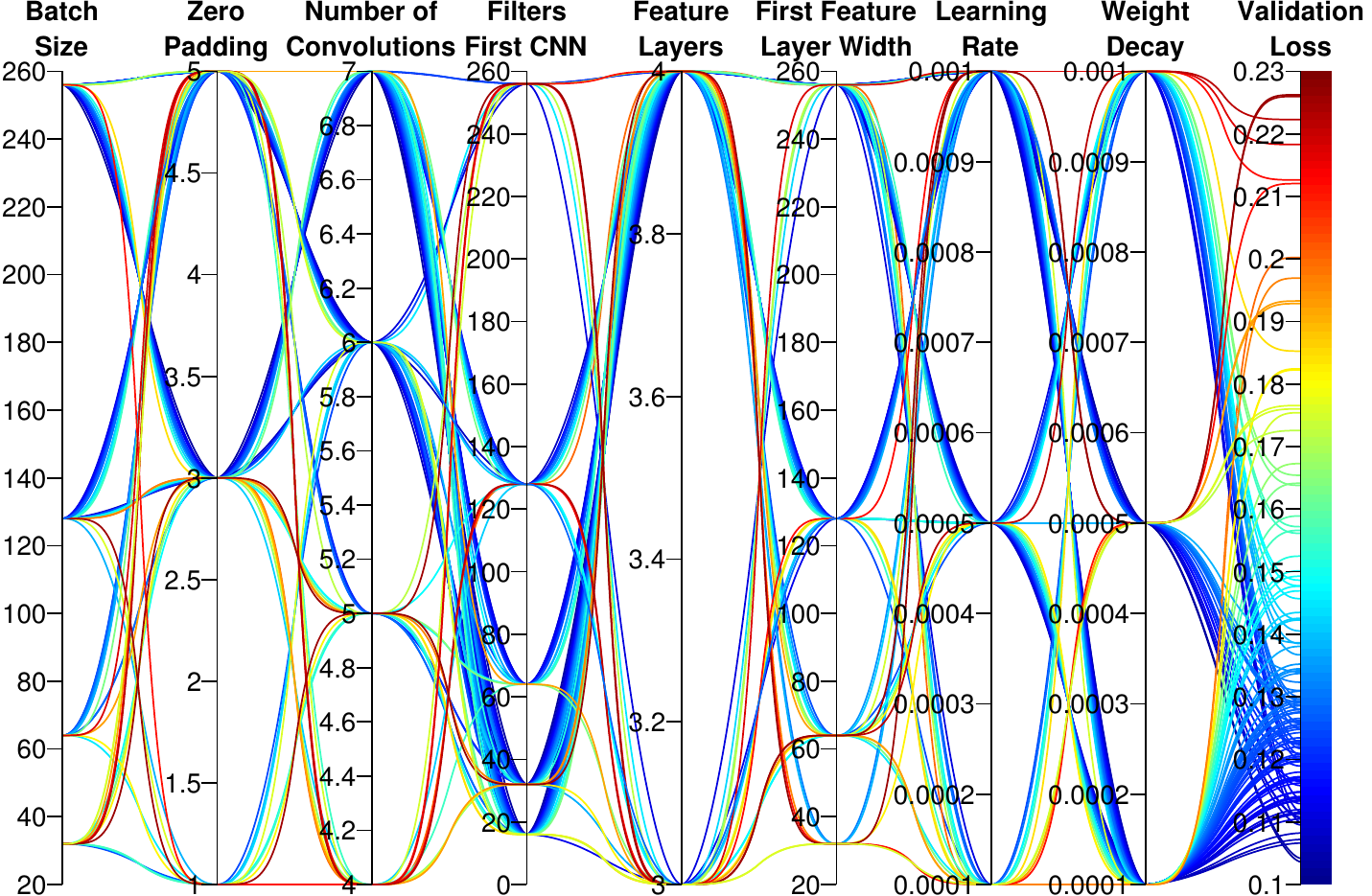}
	\vspace{-0.5cm}	
	\caption{Determination of the deep learning hyperparameter configuration using Bayesian optimization.}
	\label{fig:bayesian_optimization}
	\vspace{-0.5cm}	
\end{figure}

%
%
In the following, a sequential approach for analyzing the performance of the novel method is presented. After the initial hyperparameter optimization phase, the performance of \dragon is evaluated on the \dortmund data set and compared to the reference methods. Finally, the generalizability of the proposed approach is discussed by taking the other evaluation scenarios into account.

%
\textbf{Hyperparameter optimization:} In order to find the most suitable parameterization for \dragon, Bayesian optimization is applied using the \texttt{wandb} toolkit. Hereby, a probabilistic model of the objective function is constructed and the relationship of the different hyperparameters is learned through sequential integration of the achieved knowledge of previous training runs.  
%
%
The parallel coordinate plot in Fig.~\ref{fig:bayesian_optimization} allows to investigate the trends of the 213 investigated hyperparameter combinations. In addition, it also illustrates the sensitivity of the target variable with respect to the hyperparameter values. 
It can be observed that the most important influence factors are the number of convolutions as well as the depth and width of the feature \ac{NN}.

%
%
%
%
\begin{figure}[]  	
	\vspace{0cm}
	\centering		  
	\includegraphics[width=1.0\columnwidth]{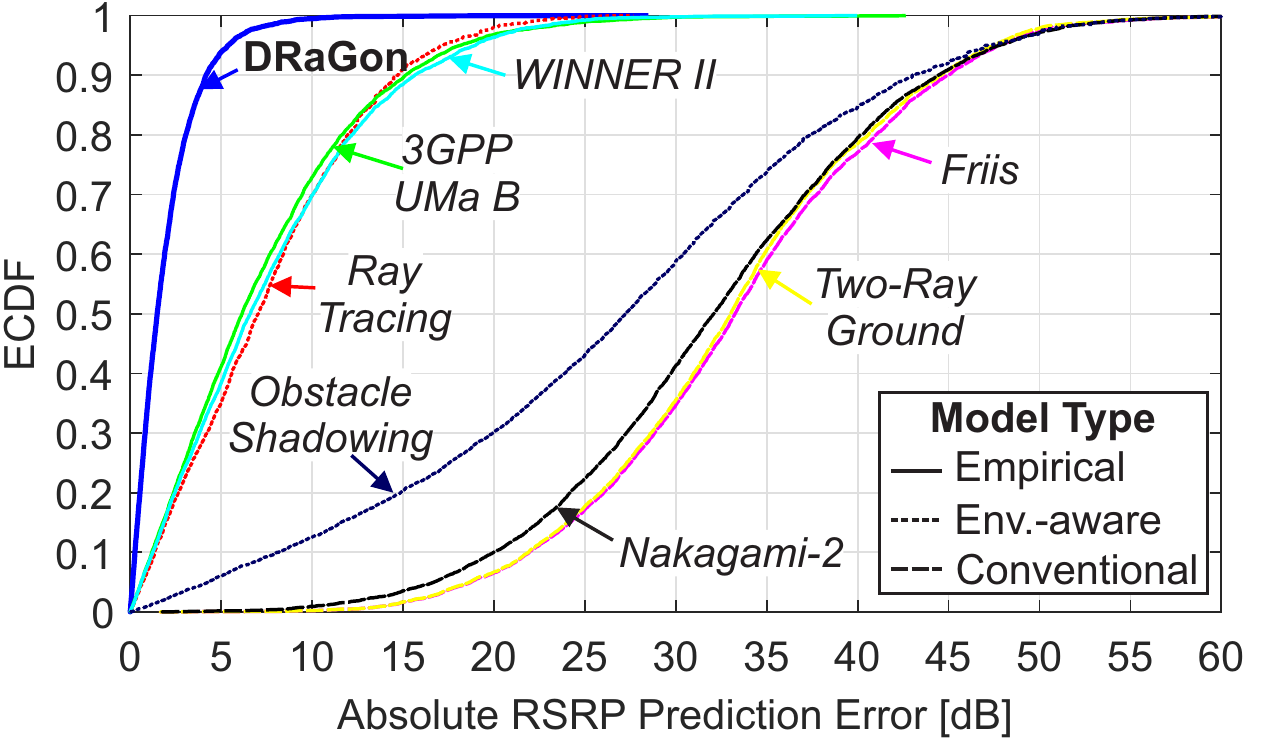}
	\vspace{-0.6cm}	
	\caption{Comparison of the absolute \ac{RSRP} prediction errors of different methods using the \dortmund data set.}
	\label{fig:dortmund_methods_ecdf}
	\vspace{-0.6cm}		
\end{figure}
%
%
\textbf{Performance comparison and validation:} The behavior of the absolute \ac{RSRP} prediction error of the considered methods is shown in Fig.~\ref{fig:dortmund_methods_ecdf} for the \dortmund data set. It can be seen that major differences rather occur between different model categories than between the individual models themselves.
%
%
While the highest prediction accuracy (\SI{2.7}{\decibel}~\ac{RMSE}) is achieved by the proposed \dragon method, the ray tracing approach shows a significantly higher prediction error (\SI{9.2}{\decibel}~\ac{RMSE}), similar to the empirical methods that achieve \SIrange{9.3}{9.6}{\decibel} \ac{RMSE}.
%
%
In compliance with \cite{Thrane/etal/2020b}, this observation shows that the performance of the ray tracing approach is significantly limited by the comparably coarse-grained \ac{OSM} data. Moreover, in our evaluations, ray tracing achieves an approximately four times lower temporal efficiency than \dragon.
%
%
It is remarked that the analytical models also make use of the transmission power estimation for $\tilde{P}_\text{TX}$ (see Sec.~\ref{sec:data_preprocessing}) and thus might be optimistic towards the \ac{3GPP} \ac{UMa} model.
%
%
The conventional channel models show a significantly lower prediction accuracy as they do not account for the obstacles within the environment. Each of the considered methods achieves at least \SI{33}{\decibel} \ac{RMSE}, which highly limits their practical applicability. 
%
%
For \dragon, it is further remarked that no significant differences of the prediction accuracy were identified for different \acp{MNO} and carrier frequency bands.

%
%
\textbf{Generalizability}: In order to analyze the generalizability of the proposed method over different scenarios, multiple data aggregation approaches are compared. 
%
%
While the \emph{scenario-wise} approach performs an individual split evaluation (see Sec.~\ref{sec:methods}) for each scenario, the \emph{global} model is trained using \SI{80}{\percent} of the aggregated data.
%
%
Finally, a \emph{cross-scenario} evaluation is performed. For each of the $i$ scenario subsets, $\mathcal{D}_{i}$ is selected as the test set $\mathcal{D}_{\text{test}}$. The remaining data sets jointly form the training set $\mathcal{D}_{\text{train}}$.
%
%
\begin{figure}[]  	
	\vspace{0cm}
	\centering		  
	\includegraphics[width=1.0\columnwidth]{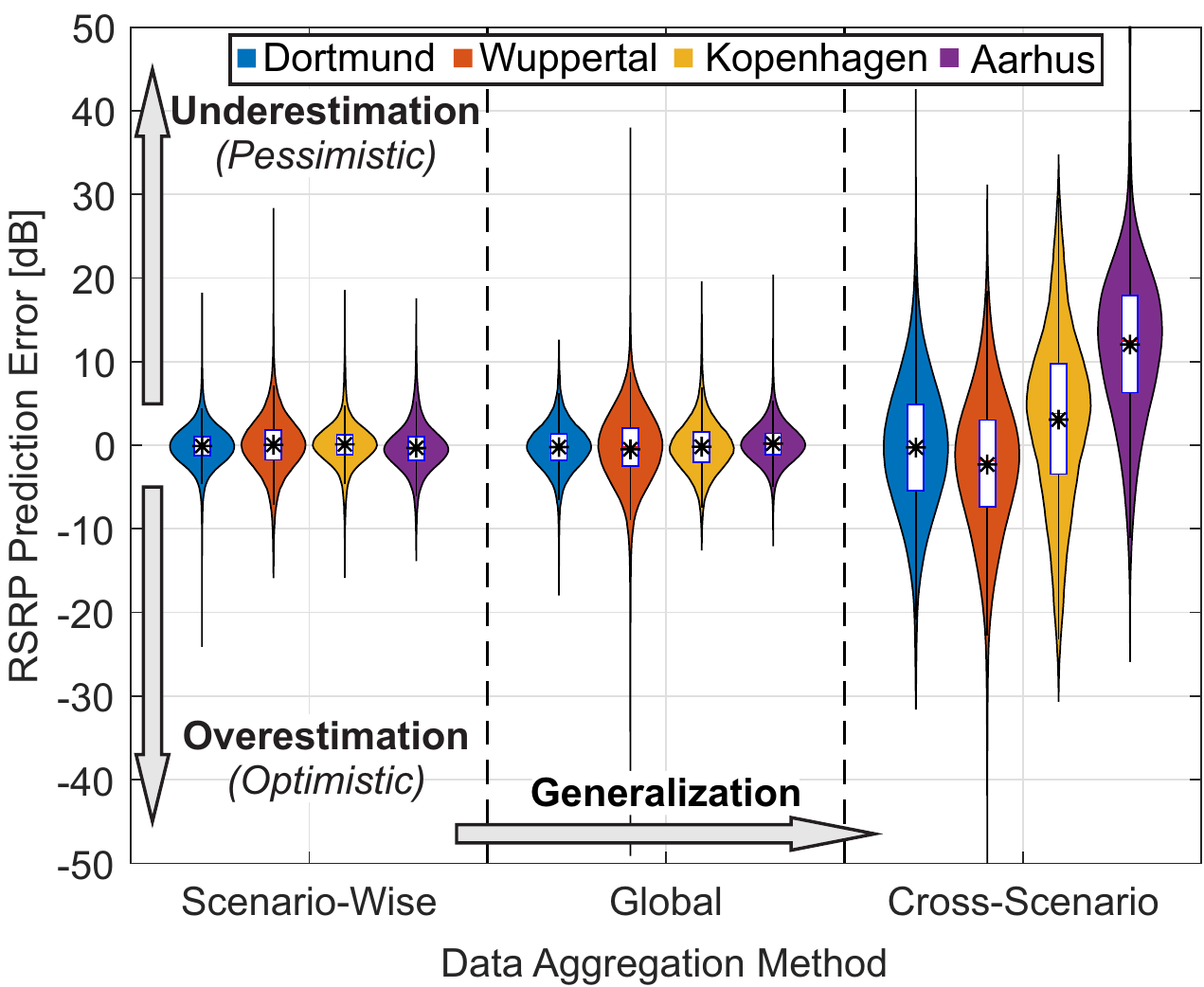}
	\vspace{-0.6cm}	
	\caption{Comparison of the scenario-specific prediction accuracy of \dragon using different data aggregation methods. In contrast to the other scenarios that utilize ground-based vehicular measurement data, the \aarhus data contains \ac{UAV} measurements at different altitudes.}
	\label{fig:cross_scenario_violin}
	\vspace{-0.7cm}	
\end{figure}
%
%
Fig.~\ref{fig:cross_scenario_violin} shows the \ac{RSRP} prediction error for the different data aggregation methods and evaluation scenarios. The consequences of prediction errors highly depend on the targeted use case: While \ac{RSRP} underestimations can lead to violations of regulation thresholds within network planning, overestimations of the network quality can lead to inefficient resource usage in anticipatory mobile networking.
%
%
Although it is well-known that the training data of machine learning methods should show a high grad of versatility for achieving good generalization \cite{Domingos/2012a}, sometimes a better scenario-specific behavior can be achieved by increasing the grade of locality of the models. \dragon shows no significant differences between those two approaches as the scenario-wise and the global data aggregation methods achieve a very similar behavior with zero mean for all evaluation scenarios. 
%
%
For the generalization, the most challenging method is the cross-scenario evaluation as it reveals the structural differences between the training data sets. 
%
%
As both German scenarios contain measurements from similar environment types (urban, suburban, rural, highway), a comparably high cross-scenario prediction accuracy is achieved for \dortmund and \wuppertal. In contrast to that, the \kopenhagen data does only contain campus measurements with moderate \ac{LOS} dynamics and low vehicle speeds. As the cross-scenario evaluation excludes this data subset from the model training, the properties of the German radio measurements are overemphasized, leading to an underestimation of the less challenging \kopenhagen environment.
%
%
The \aarhus data set exclusively contains \ac{UAV} measurements at receiver heights up to \SI{100}{\meter}. As the elevation angle increases with higher flight altitudes, the \ac{LOS} probability is also increased. However, \dragon is not able to learn this aspect if only ground-based vehicular measurements are utilized for the training. As a consequence, \dragon behaves pessimistically and underestimates the signal strength.

%
%
It is remarked that the \dortmund and \wuppertal data sets of \cite{Sliwa/Wietfeld/2019a} have been acquired using the native \texttt{Android} \ac{API} with a sampling interval of \SI{1}{\second}. As a consequence, there exists a measurement-related inaccuracy in determining the receiver location even for perfect \ac{GNSS} fixes. For the highway track with maximum driven velocity of \SI{150}{\km\per\hour}, this error can be up to \SI{42}{\meter}. Since the \dragon approach is sensitive to the quality of the geospatial information, future work should utilize more precise \ac{GNSS} sensors within the data acquisition phase.

\section{Conclusion}

%
%
In this paper, we presented the novel \dragon method for \ac{RSRP} prediction through extracting latent radio propagation information from geographical data. 
%
%
As demonstrated in a comprehensive performance evaluation campaign, the combination of expert knowledge and machine learning allows to achieve more accurate prediction results than existing methods such as analytical channel modeling and ray tracing.
It is well-known in data science literature \cite{Domingos/2012a} that often, machine learning models benefit more from additional training data than from fine-tuning of the model hyperparameters. Therefore, we aim to further increase the versatility and the quality of our training data.
%
%
In future work, we plan to utilize and further improve \dragon in the context of machine learning-enabled network planning \cite{Bektas/etal/2021a}. In addition we will consider more lightweight machine learning models \cite{Sliwa/etal/2020c} for achieving a better computational efficiency during training and inference.

\ifdoubleblind

\else

	\section*{Acknowledgment}
	
	\footnotesize
	This work has been supported by the German Research Foundation (DFG) within the Collaborative Research Center SFB 876 ``Providing Information by Resource-Constrained Analysis'', projects A4 and B4, as well as by the Ministry of Economic Affairs, Innovation, Digitalisation and Energy of the State of North Rhine-Westphalia (MWIDE NRW) along with the \emph{Competence Center 5G.NRW} under grant number 005-01903-0047 and the project \emph{Plan\&Play} under the funding reference 005-2008-0047.

\fi

\balance

\ifacm
\else
	\bibliographystyle{IEEEtran}
	\bibliography{Bibliography}

\begin{thebibliography}{10}
\providecommand{\url}[1]{#1}
\csname url@samestyle\endcsname
\providecommand{\newblock}{\relax}
\providecommand{\bibinfo}[2]{#2}
\providecommand{\BIBentrySTDinterwordspacing}{\spaceskip=0pt\relax}
\providecommand{\BIBentryALTinterwordstretchfactor}{4}
\providecommand{\BIBentryALTinterwordspacing}{\spaceskip=\fontdimen2\font plus
\BIBentryALTinterwordstretchfactor\fontdimen3\font minus
  \fontdimen4\font\relax}
\providecommand{\BIBforeignlanguage}[2]{{%
\expandafter\ifx\csname l@#1\endcsname\relax
\typeout{** WARNING: IEEEtran.bst: No hyphenation pattern has been}%
\typeout{** loaded for the language `#1'. Using the pattern for}%
\typeout{** the default language instead.}%
\else
\language=\csname l@#1\endcsname
\fi
#2}}
\providecommand{\BIBdecl}{\relax}
\BIBdecl

\bibitem{Taufique/etal/2017a}
A.~{Taufique}, M.~{Jaber}, A.~{Imran}, Z.~{Dawy}, and E.~{Yacoub}, ``Planning
  wireless cellular networks of future: {O}utlook, challenges and
  opportunities,'' \emph{IEEE Access}, vol.~5, pp. 4821--4845, 2017.

\bibitem{Cavalcanti/etal/2018a}
E.~R. Cavalcanti, J.~A.~R. de~Souza, M.~A. Spohn, R.~C. d.~M. Gomes, and A.~F.
  B. F.~d. Costa, ``{VANETs}' research over the past decade: {O}verview,
  credibility, and trends,'' \emph{SIGCOMM Comput. Commun. Rev.}, vol.~48,
  no.~2, pp. 31--39, May 2018.

\bibitem{Bui/etal/2017a}
N.~Bui, M.~Cesana, S.~A. Hosseini, Q.~Liao, I.~Malanchini, and J.~Widmer, ``A
  survey of anticipatory mobile networking: Context-based classification,
  prediction methodologies, and optimization techniques,'' \emph{IEEE
  Communications Surveys \& Tutorials}, 2017.

\bibitem{Sliwa/etal/2021b}
B.~Sliwa, R.~Adam, and C.~Wietfeld, ``Client-based intelligence for resource
  efficient vehicular big data transfer in future {6G} networks,'' \emph{IEEE
  Transactions on Vehicular Technology}, Feb 2021.

\bibitem{Yun/Iskander/2015a}
Z.~Yun and M.~F. Iskander, ``Ray tracing for radio propagation modeling:
  {P}rinciples and applications,'' \emph{IEEE Access}, vol.~3, pp. 1089--1100,
  2015.

\bibitem{Thrane/etal/2020b}
J.~{Thrane}, D.~{Zibar}, and H.~L. {Christiansen}, ``Model-aided deep learning
  method for path loss prediction in mobile communication systems at {2.6
  GHz},'' \emph{IEEE Access}, vol.~8, pp. 7925--7936, 2020.

\bibitem{Ali/etal/2020a}
S.~Ali, W.~Saad, N.~Rajatheva, K.~Chang, D.~Steinbach, B.~Sliwa, C.~Wietfeld,
  K.~Mei, H.~Shiri, H.-J. Zepernick, T.~M.~C. Chu, I.~Ahmad, J.~Huusko,
  J.~Suutala, S.~Bhadauria, V.~Bhatia, R.~Mitra, S.~Amuru, R.~Abbas, B.~Shao,
  M.~Capobianco, G.~Yu, M.~Claes, T.~Karvonen, M.~Chen, M.~Girnyk, and
  H.~Malik, ``{6G} white paper on machine learning in wireless communication
  networks,'' Apr 2020.

\bibitem{Sliwa/Wietfeld/2019a}
B.~Sliwa and C.~Wietfeld, ``Data-driven network simulation for performance
  analysis of anticipatory vehicular communication systems,'' \emph{IEEE
  Access}, Nov 2019.

\bibitem{Wang/etal/2020b}
C.~{Wang}, J.~{Huang}, H.~{Wang}, X.~{Gao}, X.~{You}, and Y.~{Hao}, ``6g
  wireless channel measurements and models: {T}rends and challenges,''
  \emph{IEEE Vehicular Technology Magazine}, vol.~15, no.~4, pp. 22--32, 2020.

\bibitem{Goodfellow/etal/2016a}
I.~Goodfellow, Y.~Bengio, and A.~Courville, \emph{Deep Learning}.\hskip 1em
  plus 0.5em minus 0.4em\relax MIT Press, 2016,
  \url{http://www.deeplearningbook.org}.

\bibitem{5GAA/2020a}
5GAA, ``White {p}aper: {M}aking {5G} proactive and predictive for the
  automotive industry,'' 5G Automotive Association, Tech. Rep., Jan 2020.

\bibitem{Poegel/Wolf/2015a}
T.~{Pögel} and L.~{Wolf}, ``Optimization of vehicular applications and
  communication properties with connectivity maps,'' in \emph{2015 IEEE 40th
  Local Computer Networks Conference Workshops (LCN Workshops)}, Oct 2015, pp.
  870--877.

\bibitem{Kliks/etal/2020a}
A.~{Kliks}, L.~{Kulacz}, P.~{Kryszkiewicz}, H.~{Bogucka}, M.~{Dryjanski},
  M.~{Isaksson}, G.~P. {Koudouridis}, and P.~{Tengkvist}, ``Beyond {5G}: {B}ig
  data processing for better spectrum utilization,'' \emph{IEEE Vehicular
  Technology Magazine}, vol.~15, no.~3, pp. 40--50, 2020.

\bibitem{Viriyasitavat/etal/2015a}
W.~{Viriyasitavat}, M.~{Boban}, H.~{Tsai}, and A.~{Vasilakos}, ``Vehicular
  communications: {S}urvey and challenges of channel and propagation models,''
  \emph{IEEE Vehicular Technology Magazine}, vol.~10, no.~2, pp. 55--66, 2015.

\bibitem{3GPP/2019d}
``{3GPP TR 38.901} - {S}tudy on channel model for frequencies from 0.5 to 100
  {GHz}, {V 16.1.0},'' {3rd Generation Partnership Project (3GPP)}, Tech. Rep.
  38.901, Dec 2019.

\bibitem{Sommer/etal/2014a}
C.~{Sommer}, D.~{Eckhoff}, and F.~{Dressler}, ``{IVC} in cities: {S}ignal
  attenuation by buildings and how parked cars can improve the situation,''
  \emph{IEEE Transactions on Mobile Computing}, vol.~13, no.~8, pp. 1733--1745,
  2014.

\bibitem{Wang/etal/2020a}
J.~{Wang}, C.~{Jiang}, H.~{Zhang}, Y.~{Ren}, K.~{Chen}, and L.~{Hanzo},
  ``Thirty years of machine learning: {T}he road to pareto-optimal wireless
  networks,'' \emph{IEEE Communications Surveys Tutorials}, pp. 1--1, 2020.

\bibitem{Enami/etal/2018a}
R.~Enami, D.~Rajan, and J.~Camp, ``{RAIK}: {R}egional analysis with geodata and
  crowdsourcing to infer key performance indicators,'' in \emph{2018 IEEE
  Wireless Communications and Networking Conference (WCNC)}, April 2018, pp.
  1--6.

\bibitem{Masood/etal/2019a}
U.~{Masood}, H.~{Farooq}, and A.~{Imran}, ``A machine learning based {3D}
  propagation model for intelligent future cellular networks,'' in \emph{2019
  IEEE Global Communications Conference (GLOBECOM)}, Dec 2019, pp. 1--6.

\bibitem{MorochoCayamcela/etal/2020a}
M.~E. {Morocho-Cayamcela}, M.~{Maier}, and W.~{Lim}, ``Breaking wireless
  propagation environmental uncertainty with deep learning,'' \emph{IEEE
  Transactions on Wireless Communications}, vol.~19, no.~8, pp. 5075--5087,
  2020.

\bibitem{Thrane/etal/2020a}
J.~Thrane, B.~Sliwa, C.~Wietfeld, and H.~Christiansen, ``Deep learning-based
  signal strength prediction using geographical images and expert knowledge,''
  in \emph{2020 IEEE Global Communications Conference (GLOBECOM)}, Taipei,
  Taiwan, Dec 2020.

\bibitem{3GPP/2020a}
``{3GPP TS 38.215} - {P}hysical layer measurements, {V 16.4.0},'' {3rd
  Generation Partnership Project (3GPP)}, Tech. Rep., Dec 2020.

\bibitem{Bashfield/Keim/2011a}
A.~Bashfield and A.~Keim, ``Continent-wide {DEM} creation for the {European
  Union},'' in \emph{34th International Symposium on Remote Sensing of
  Environment}, Apr. 2011.

\bibitem{Sliwa/etal/2019c}
B.~Sliwa, M.~Patchou, and C.~Wietfeld, ``Lightweight simulation of hybrid
  aerial- and ground-based vehicular communication networks,'' in \emph{2019
  IEEE 90th Vehicular Technology Conference (VTC-Fall)}, Honolulu, Hawaii, USA,
  Sep 2019.

\bibitem{Lopez/etal/2019a}
M.~{Lopez}, T.~B. {Sorensen}, P.~{Mogensen}, J.~{Wigard}, and I.~Z. {Kovacs},
  ``Shadow fading spatial correlation analysis for aerial vehicles: {R}ay
  tracing vs. measurements,'' in \emph{2019 IEEE 90th Vehicular Technology
  Conference (VTC2019-Fall)}, 2019, pp. 1--5.

\bibitem{Sommer/etal/2011a}
C.~{Sommer}, D.~{Eckhoff}, R.~{German}, and F.~{Dressler}, ``A computationally
  inexpensive empirical model of {IEEE} 802.11p radio shadowing in urban
  environments,'' in \emph{2011 Eighth International Conference on Wireless
  On-Demand Network Systems and Services}, 2011, pp. 84--90.

\bibitem{Domingos/2012a}
P.~Domingos, ``A few useful things to know about machine learning,''
  \emph{Commun. ACM}, vol.~55, no.~10, p. 78–87, Oct. 2012.

\bibitem{Bektas/etal/2021a}
C.~Bektas, S.~B{\"o}cker, B.~Sliwa, and C.~Wietfeld, ``Rapid network planning
  of temporary private {5G} networks with unsupervised machine learning,'' in
  \emph{2021 IEEE 94rd Vehicular Technology Conference (VTC-Fall)}, Virtual,
  Sep 2021.

\bibitem{Sliwa/etal/2020c}
B.~Sliwa, N.~Piatkowski, and C.~Wietfeld, ``{LIMITS}: {L}ightweight machine
  learning for {IoT} systems with resource limitations,'' in \emph{2020 IEEE
  International Conference on Communications (ICC)}, Dublin, Ireland, Jun 2020,
  {B}est paper award.

\end{thebibliography}
\fi

\end{document}